\documentclass[showpacs,twocolumn,prl]{revtex4}
\usepackage{graphicx}
\usepackage{epsfig}
\usepackage{amssymb}

\def\beq{\begin{equation}}
\def\eeq{\end{equation}}
\def\beqa{\begin{eqnarray}}
\def\eeqa{\end{eqnarray}}

\def\e{\epsilon}

\def\half{{\ss 1\over 2}}

\def\D{\Delta}
\def\del{\delta}
\def\e{\epsilon}
\def\cH{{\mathcal H}}

\def\ss{\scriptstyle}
\def\hcon{\mathrm{H.c.}}

\def\R{\mathrm{R}}
\def\L{\mathrm{L}}

\def\ss{\scriptstyle}

\def\etal{{\sl et al.}}

\def\nonum{\nonumber \\}

\def\del{\delta}

\def\nonum{ \nonumber \\}
\renewcommand{\sec}[1]{\vskip 0.truecm \noindent \emph{#1}. -- }
\newcommand{\ket}[1]{| #1 \rangle}
\newcommand{\bra}[1]{\langle #1 |}
\newcommand{\av}[1]{\langle #1 \rangle}

\begin{document}

\title{Fourier's Law: insight from a simple derivation}
\author{Y. Dubi and M. Di Ventra}
\affiliation{Department of Physics, University of California San
Diego, La Jolla, California 92093-0319, USA}
\begin{abstract}

The onset of Fourier's law in a one-dimensional quantum system is
addressed via a simple model of weakly coupled quantum systems in
contact with thermal baths at their edges. Using analytical
arguments we show that the crossover from the ballistic (invalid
Fourier's law) to diffusive (valid Fourier's law) regimes is
characterized by a thermal length-scale, which is directly related
to the profile of the local temperature. In the same vein, dephasing
is shown to give rise to a classical Fourier's law, similarly to the
onset of Ohm's law in mesoscopic conductors.

\end{abstract}
\maketitle

\sec{Introduction} Fourier's law of heat conduction \cite{Fourier} states that when a system is subject to a temperature
difference, a uniform temperature gradient $\nabla T({\bf r})$ ensues in its interior and the heat current density $\bf j({\bf
r})$ is proportional to that gradient, $ {\bf j}({\bf r})=-k \nabla T({\bf r})$. The proportionality constant $k$ is called
thermal conductivity and could be position dependent.

Although this law is almost two centuries old, a general demonstration, or a set of conditions for its validity in the general
case, is still lacking \cite{Bonneto,Buchanan}. The problem has recently received renewed attention, both theoretically
\cite{FourierReview, ourpaper} and experimentally \cite{Chang}, especially in quantum systems, due to the growing interest in
energy transport at the nanoscale, whose understanding is an important step towards utilizing nanostructures for potential energy
applications \cite{Chang1,Chang2,Chang3}.

The difficulty in proving this law from first principles is twofold.
For one, a local temperature (in an inherently out-of-equilibrium
situation) needs to be defined. Secondly, a local heat current needs
to be calculated and these two quantities need to be compared. Both
of these quantities are difficult to evaluate for the simple reason
that they are usually neither well defined (like the case of the
local temperature \cite{LocalT}) nor their definition is unique
(for the case of the heat current \cite{Segal1}). In the quantum
case, an additional complication arises which is related to the fact that the size
of the Hilbert space of a given system generally increases
exponentially with system size \cite{Pershin}, so that the problem easily
becomes computationally very demanding.

In recent years, several models have been put forward where Fourier's law has been demonstrated
\cite{Lepri,Garrido,Saito,Michel,Li,Bonetto2,Bricmont,Gaspard,Dhar,Gaul,Roy,Lebowitz,Roy2,Jacquet}, all of which employ in some
way the idea of local equilibrium. The local temperature is usually either calculated from the expectation value of some local
energy operator \cite{Saito,Michel2,Mejia,Gaul}, or a uniform temperature gradient is assumed to exist \cite{Segal1}. An
alternative route is to study the energy diffusion in closed systems (i.e. without thermal baths) \cite{Steinigeweg}, or to study
a system with self-consistent reservoirs \cite{Lebowitz,Dhar,Roy2,Jacquet}.

In Ref.~\cite{ourpaper}, the present authors utilized a novel scheme
from the theory of open quantum systems \cite{Pershin} to calculate
the local temperature and the onset of Fourier's law in electronic
quantum wires. The main findings were that for a ballistic system
Fourier's law is invalid, and the temperature is constant along the
wire. A temperature gradient develops as disorder is introduced in
the wire, eventually leading to the onset of Fourier's law. These
results, however, relied on numerical simulations and no simple
analytical form could be deduced.

In this paper we aim at calculating analytically both the local
temperature and the heat current for a model one-dimensional quantum
system. Our results indicate that Fourier's law is related to a
thermal length scale, either quantum (localization length) or
classical (dephasing length), which is reflected in the local
temperature profile. Fourier's law is then valid only if this
thermal scale is smaller than the system length. In particular,
dephasing gives rise to a classical Fourier's law, similarly to the
onset of Ohm's law in mesoscopic conductors.

\sec{Model} We consider a set of $N$ weakly-connected identical sub-systems, $S_n$. Each sub-system $S_n$ has its own Hamiltonian
$\cH_n=\sum_k \e_k \ket{k^{(n)}}\bra{k^{(n)}}$, where $\ket{k^{(n)}}$ are the many-body state vectors of $S_n$. The sub-systems
are connected via a tunneling Hamiltonian $\cH_{n,n+1}=\sum_{k,k'} V_{kk'}\ket{k^{(n)}}\bra{k^{(n+1)}}+\hcon$. The full
Hamiltonian is then $\cH=\sum_n (\cH_n+\cH_{n,n+1})$.

\begin{figure}[h!]
\vskip 0.5truecm
\includegraphics[width=7truecm]{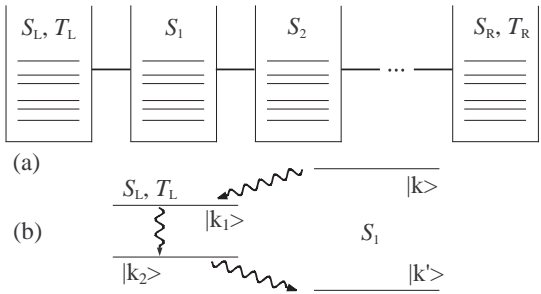}
\caption{(a) Schematic representation of the model. The system consists of identical, weakly connected sub-systems. The left-and
right-most sub-systems are held at temperatures $T_\L$ and $T_\R$, respectively. (b) Representation of the transition process in
$S_1$ (see text).}\label{fig1}
\end{figure}

Here we assume that only the left-most and right-most sub-systems ($S_\L$ and $S_\R$) are connected to external environments, and
are consequently held at temperatures $T_\L$ and $T_\R$, respectively. The other sub-systems are not coupled to an external
environment, and hence transitions between states in the central sub-systems have to be mediated by the edge sub-systems. The
model is schematically depicted in Fig. ~\ref{fig1}.

\sec{Local temperature} The aim of the calculation is to evaluate the local temperature of $S_n$ as a function of $T_\L, ~T_\R$
and $n$ -- which serves as a position variable. The main \emph{assumption} of the model is that the tunneling interaction is weak
enough such that each sub-system is in a local thermal equilibrium at a certain temperature $T_n$, described below.

Let us start by considering only $\cH_\L$ and a single additional sub-system $\cH_1$. By stating that $S_\L$ is maintained at a
temperature $T_\L$, we assume that it is in a local thermal equilibrium. Due to the external environment (which determines the
temperature), there are transitions between different states $k$ and $k'$, which are defined via the scattering rates
$W^{(\L)}_{k\to k'}$. The statistical meaning of the temperature is that there is a single number $T_\L$, which characterizes all
the different transitions by a single rule of detailed balance. This means that regardless of $k$ and $k'$, the transition rates
obey $\frac{W^{(\L)}_{k\to k'}}{W^{(\L)}_{k'\to k}}=\exp\left(-\frac{\D \e_{kk'}}{T_\L} \right)$ (taking $k_B=\hbar=1$ throughout
the paper), where $\D \e_{kk'}=\e_{k'}-\e_{k}$.

Now consider the system $S_1$. Since $S_1$ is described in the
energy basis, a transition between states in $S_1$ is an inelastic
process, which (in the absence of interactions) requires the
presence of an environment. Since the only subsystem in contact with
an environment is $S_L$, for a particle in a state $k$ in $S_1$ to
scatter to a different state $k'$ in $S_1$, a scattering event to
some state $k_1$ in $S_\L$ has to first occur. From that state,
another transition will occur to a state $k_2$ in $S_\L$ and a final
transition to $k'$ (this process is depicted in Fig.1(b)). Thus, the
transition rate in $S_1$ is given by \beq W^{(1)}_{k\to
k'}=\sum_{k_1,k_2}\Gamma^{1\to \L}_{k\to k_1} W^{(\L)}_{k_1 \to k_2}
\Gamma^{\L \to 1}_{k_2\to k'} ~~,\label{rate1} \eeq where
$\Gamma^{n\to n+1}_{k\to k'}$ is a transition probability from the
state $k$ in $n$ to the state $k'$ in $n'$, and is therefore
proportional to the overlap between $\ket{k^{(n)}}$ and
$\ket{k'^{(n')}}$, and hence to $|V_{kk'}|^2$ (in similarity to the
Fermi golden rule). We now assume that the main contribution comes
from states of the same energy, and for simplicity take a uniform
tunneling Hamiltonian, i.e. $V_{kk}=V$, which yields $\Gamma^{n\to
n+1}_{k\to k'}=\Gamma \del_{kk'}$. This implies \beq W^{(1)}_{k\to
k'}=\gamma W^{(\L)}_{k \to k'} ~~, \label{rate2} \eeq where
$\gamma=\Gamma^2$ (note that in our notation states labeld by $k$
may belong to different subsystems, denoted by the upper index
$(n)$). Now, the temperature of $S_1$ may be obtained from the
detailed balance of $S_1$, via $\frac{W^{(1)}_{k\to
k'}}{W^{(1)}_{k'\to k}}=\exp\left(-\frac{\D \e_{kk'}}{T_1} \right)$.
Employing Eq.~(\ref{rate1}), the $\gamma$ prefactor cancels and we
find that $T_1=T_L$. This is a simple manifestation of the fact that
two systems in contact with each other equilibrate.

The next step is to consider a chain of $N$ sub-systems, still only connected to a single $S_L$. The goal now is to find the
temperature of $S_n$. One can repeat the procedure described above, with the only change being that $n$ intermediate, neighboring
system transitions occur before the particles have a transition event at $S_L$. Therefore, a simple generalization gives $
W^{(n)}_{k\to k'}=\gamma^n W^{(\L)}_{k \to k'} \label{rate3} $, which again gives $T_n=T_\L$. This argument is valid to lowest
order in $\gamma$, which physically corresponds to including only sequential tunneling processes.

By considering the addition of the right-most system $S_\R$ and the
edge of the $N$-long chain, one similarly obtains \beq W^{(n)}_{k\to
k'}=\gamma^n W^{(\L)}_{k \to k'}+\gamma^{N-n} W^{(\R)}_{k \to k'}
\label{rate3} ~~.\eeq In this case, due to the presence of the right
temperature $T_R$ one cannot simply cancel out the prefactor, and
the expression for $T_L$ becomes more complex. In order to make
progress, we assume that $T_\R=T_\L+\del T,~~\del T << T_L$, and
assume that the transition rates take the form $W^{(\L,\R)}_{k \to
k'} \propto \exp \left( -\frac{\D \e_{kk'}}{T_{\L,\R}}\right)$ (in
agreement with the assumption that $S_L$ and $S_R$ are at
equilibrium). Substituting back into Eq.~(\ref{rate3}) and taking
the first order in $\del T$, we find (after some algebraic
manipulation) \beq T_n=T_L+\del T \frac{1}{1+\gamma^{2n-N}} ~~.
\label{localT1}\eeq Defining a position variable $x=\frac{2n}{N}-1$
we obtain $T(x)=\frac{1}{1+\exp(x/\xi)}$, where $\xi=(N \log
\gamma)^{-1}$ is the ``thermal length'' that defines the
length-scale over which there is an (approximately) uniform
temperature gradient.

\begin{figure}[h!]
\vskip 0.5truecm
\includegraphics[width=7truecm]{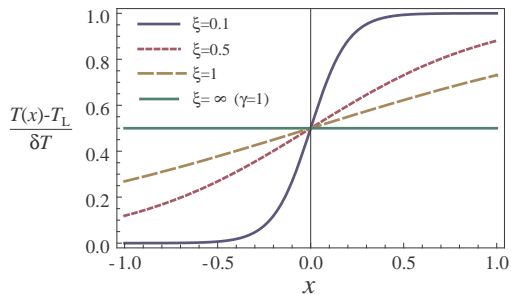}
\caption{The (normalized) local temperature $(T(x)-T_\L)/\del T$ as a function of the position variable $x$. This figure should
be compared with Fig.~1 of Ref.~\cite{ourpaper}.}\label{fig2}
\end{figure}

In Fig.~\ref{fig2} we plot $(T(x)-T_\L)/\del T$ (which scales from $0$ to $1$) as a function of position variable $x$, for
different values of $\xi=0.1,0.5,1$ and $\xi=\infty$, which corresponds to $\gamma=1$. For small values of $\gamma$ (and hence of
$\xi$), most of the temperature change is close to the middle of the wire. In those regions, a uniform temperature gradient is
indeed developed, and hence Fourier's conjecture is valid. When $\gamma=1$ the temperature is uniform in the wire,
 and hence no temperature gradient ensues. This figure should be compared with Fig.~1 of Ref.~\cite{ourpaper}, which
 exhibit similar features, albeit obtained from a microscopic model (with no {\it a priori} assumptions).

We note that while strictly speaking the case $\gamma=1$ is beyond the above perturbation analysis, this result is still valid.
This is because in the $\gamma=1$ case all the wave functions are delocalized and span the entire system. Thus, $S_\L$ and $S_\R$
have the same weight in the transition rates of $S_n$ regardless of $n$, giving rise to a uniform temperature (which is just
$T_n=T_L+\half \del T$ for a small temperature difference).

\sec{Heat current} The next step towards understanding Fourier's law
in our model system is to calculate the heat current. If the
sub-systems are weakly coupled to each other, the local energy is
naturally defined~\cite{Segal1} as $E_n=\langle \cH_n \rangle$.
From the continuity equation \cite{Segal1,ourpaper} one has a simple
expression for the heat current, $j_n=-\Gamma (E_{n+1}-E_{n})$
(taking the distance between the sub-systems to be $a=1$). This
definition for the heat current could be understood by noting that
(under the assumption of a constant $\Gamma$) $\Gamma E_{n+1}$ is
simply the rate of energy flow from $S_{n+1}$ to $S_n$, and vice
versa, and thus $j_n$ defined above describes the net energy flow
between $S_{n+1}$ and $S_n$ per unit time, i.e., the heat current.

To calculate $j_n$, we assume that for each sub-system $S_n$ we can
define the probability $P^{(n)}_k$ to find the system in the state $
\ket{k^{(n)}} $. Then, we have $E_n=\sum_k \e_k P^{(n)}_k $, and for
the heat current \beqa j_n & =& -\Gamma \sum_k \e_k
(P^{(n+1)}_k-P^{(n)}_k ) \nonum & \approx & -\Gamma \sum_k
\e_k\frac{\partial P^{(n)}_k}{\partial n} = -\Gamma \sum_k\e_k
\frac{\partial P^{(n)}_k}{\partial T_n} \frac{\partial T_n}{\partial
n} ~~. \label{J1} \eeqa Since we are assuming local thermal
equilibrium, it follows that $P^{(n)}_k \propto \exp(-\e_k/T_n)$,
and hence $\frac{\partial P^{(n)}_k}{\partial
T_n}=\frac{\e_k}{T^2_n}P^{(n)}_k$. This gives for the heat current
\beq j_n=-\Gamma \sum_k \left( \frac{\e^2_k P^{(n)}_k}{T^2_n}\right)
\nabla T(n)=-\kappa_n \nabla T(n)~~,\label{J2}\eeq where
$\kappa=\frac{\Gamma \av{\e^2}}{T^2_n}$, in agreement with the
standard expectations \cite{Peierls,Segal2}. Note that $j$ is
proportional to the temperature gradient, and hence the thermal
conductance scales as $L^{-1}$, as required by Fourier's law.
However, since the local temperature itself has a length-dependence,
the above rule strictly applies only within the thermal length
$\xi/2$ from the center of the wire. Since in experiments the
measured $\kappa$ is a global property (that is, an average of
$\kappa$ over the entire length of the sample) finite size effects
may take place and give unusual scaling for $\kappa(L)$
\cite{Chang}.

The fact that this simple model exhibits Fourier's law at weak coupling can also be understood by comparing it to the results of
Ref.~\cite{Segal1}, in which a general form of Fourier's law is derived for a general system which obeys three conditions. Our
model satisfies these conditions, and hence one indeed expect it to display Fourier's law.

However, we stress that the existence of a temperature gradient does
not necessarily imply ``normal'' heat conduction. While in a simple
system as that described here we have shown that the two are
connected, in more complicated systems, the non-linear nature of the
interactions may give rise to a finite temperature gradient but
anomalous conductance \cite{Hu}.

\sec{Dephasing} Up to now we have assumed that there is contact between the chain and the external environment only at the edges
of the chain. Let us discuss the effect of local environments acting along the chain, which may result in the dephasing of the
wave-functions (in real systems this may be caused by, e.g. inelastic scattering off low-energy phonons). We thus introduce a
length scale $L_{\phi}$, which characterizes the length over which the wave function retains its phase. If $L_\phi>N$ then the
dephasing has no effect, and Fourier's law is valid in its form of Eq.~(\ref{localT1}). However, in the case $L_\phi<N$, there is
a new natural division of the system into subsystems of length $L_\phi$.~\cite{DiVentrabook} Again, we assume that each of these
subsystems is in local equilibrium, and interacts weakly with the neighboring subsystem. However, as opposed to the case
discussed above, there is no coherent tunneling between the subsystems, but rather classical transport between them. Thus, the
rate equations for the occupation probabilities in the subsystems may be written as \cite{Gurvitz}

 \beqa
 \dot{P}^{(n)}_k &=& \Gamma_\phi(P^{(n+1)}_{k}-P^{(n)}_k)+\Gamma_\phi(P^{(n-1)}_{k}-P^{(n)}_k) ~~,\label{rate_dephasing}\eeqa
where, for simplicity, we have assumed that the transitions are
between states of similar energies. Here $n=1,2,... N_{\phi}$ is the
index of the sub-system, and the constant $\Gamma_\phi$ describes a
typical transition rate between sub-systems defined via the division
to $L_\phi$ sub-systems, and should in principal be determined
microscopically.

Equation~(\ref{rate_dephasing}), along with the boundary conditions
$P^{(\L,\R)}_k=P^{(\L,\R)}_{k,\mathrm{eqilibrium}} \propto
\exp(-\e_k/T_{\L,\R}) )$, has a simple solution for its steady
state, given by $P^{(n)}_k=P^{(\L)}_k+\frac{n}{N}
(P^{(\R)}_k-P^{(\L)}_k)$. Assuming that $P^{(n)}_k \propto
\exp(-\e_k/T_{n})$ and in the limit $T_\R-T_\L=\del T<<T_\L$ one
obtains a linear form for the local temperature, $T_n \approx
T_\L+\frac{n}{N} \del T$. We thus conclude that {\it dephasing
brings about the classical form of Fourier's law}. Explicit examples
can be found in, e.g., Ref.~\cite{Dhar} and Ref.~\cite{Jacquet},
where local dephasing  was introduced (in a quantum harmonic lattice
and electronic system, respectively) by means of local external
baths, giving rise to Fourier's law and a linear temperature
profile.

It is also useful to consider the analogy between the effect of
dephasing on the local temperature and on the resistance of a
one-dimensional wire consisting of localized sub-systems (i.e. an
Anderson insulator)~\cite{dephasing}. In the absence of dephasing
($L_\phi>N$) the resistance is exponential in the wire length,
$R\sim \exp(N/\xi_\mathrm{loc})$, where $\xi_\mathrm{loc}$ is the
localization length. However, when $L_\phi<N$, the resistances of
different subsystems of length $\L_\phi$ (each with a resistance of
$R_\phi \sim \exp(L_\phi/\xi_\mathrm{loc})$) are connected in
series, resulting in a linear dependence of the resistance on
length, $R \sim R_\phi L/L_\phi$, i.e., classical Ohm's law. This
crossover from a classical to a quantum regime is similar to what
was demonstrated above in the case of Fourier's law.

 \sec{Discussion} In summary, we have presented a simple model where the local temperature and heat
currents may be evaluated analytically. We have shown that the onset
of Fourier's law requires the presence of a local thermal
equilibrium at each sub-system that constitutes the full system. It
breaks down in the case of strong coupling between the sub-systems.
In that case, the temperature is constant throughout the sample.
Including dephasing processes brings about a classical form of
Fourier's law, in similarity to the onset of Ohm's law for the
resistance.

The generality of these results may be understood by the following arguments. Consider a one-dimensional system held at a
temperature difference $\del T$. Now, the system may be broken into the smallest possible sub-systems ($S_n$ in the above
calculation). If each such sub-system has a unique temperature which defines the relaxation rates between all the states in the
sub-system, then the system is in local thermal equilibrium and Fourier's law is valid. If not, the system may be coarse-grained
to generate larger sub-systems, and again for each sub-system a local temperature is sought. If local thermal equilibrium is
eventually obtained, then the number of coarse-grained sub-systems describes the effective length of the system, and along with
the interaction between the sub-systems, it gives the thermal length $\xi$ which determines the length-scale for a uniform
temperature gradient to ensue. However, if the coarse-graining procedure reaches the scale of the system, only a single
temperature can be defined. In that case, the temperature is uniform across the sample, the thermal conductivity diverges and
Fourier's law is invalid. A similar phenomenon occurs in the presence of dephasing, where the role of the (quantum) thermal
length is played by the dephasing length.

Recently, a crossover from ballistic to diffusive thermal transport
was shown numerically to appear in a system with self-consistent
reservoirs \cite{Roy2}. The crossover was determined from a
length-scale related to the coupling between the system and the
baths, in a similar fashion to the dephasing length $L_{\phi}$
discussed above. Our results are in agreement with (and thus provide
an intuitive explanation for) the numerical observations in Ref.
~\onlinecite{Roy2}. However, there are indications that in certain
models there is no length-dependent crossover \cite{Hu}. Therefore,
whether the onset of Fourier's law is always a crossover phenomenon
seems to depend on the specific model and a satisfactory answer to
this question has yet to be found.

This work has been supported by DOE under grant DE-FG02-05ER46204
and UC Labs.

\end{document}